\documentclass[final,5p,times,twocolumn]{elsarticle}

\usepackage{amsmath,amssymb}
\usepackage{bm}
\usepackage{hyperref}

\usepackage[normalem]{ulem}  
\ifpdf
  \usepackage[pdftex]{color}
\else
  \usepackage[dvipdfmx]{color} 
\fi

\begin{document}

\begin{frontmatter}



\title{The gravitational wave and short gamma-ray burst GW170817/SHB170817A,\\
not your everyday binary neutron star merger}

\author[label3]{A. De R\'ujula\\
\vspace{.3 cm}
Instituto de F\'isica Te\'orica (UAM/CSIC), Univ. Aut\'onoma de Madrid, Spain;\\
Theory Division, CERN, CH 1211 Geneva 23, Switzerland}
\ead{alvaro.derujula@cern.ch}


\begin{abstract}

This event, so far unique, beautifully  confirmed the standard views on
the gravitational waves produced by a merger of two neutron stars, but 
its electromagnetic multi-wavelenth observations disagreed with the
numerous initial versions of the
``standard fireball model(s)" of gamma ray bursts. Contrariwise, they provided strong
evidence in favour of the ``cannonball" model. Most uncontroversially, 
a cannonball was observed at radio wavelengths, with an overwhelming
statistical significance ($>\! 17\,\sigma$), and travelling in the plane
of the sky, as expected, at an apparent superluminal velocity $V_{app}\sim 4\, c$.

\end{abstract}


\end{frontmatter}



\section{Introduction}

The GW170817 event was the first binary neutron-star merger 
detected with Ligo-Virgo \cite{Abbott} in 
gravitational waves (GWs). It was followed by SHB170817A\footnote{SHB stands for 
``Short Hard Burst", a sub-class of gamma-ray bursts (GRBs) lasting less 
that $\sim 2$ s and whose photons generally have comparatively large energies.}, 
$1.74\!\pm\! 0.05$ s  after the end
of the GW's detection. The SHB's afterglow across the electromagnetic spectrum was
used to localize its source \cite{Hjorthetal2017} to the galaxy NGC 4993, at a cosmologically
very modest redshift, $z\!=\!0.009783$.
The GW170817/SHB170817A association was the first indisputable confirmation that pairs of
neutron stars merging due to GW emission produce GRBs, 
thereafter converting this suggestion \cite{Goodman2} \cite{Meszaros} 
into a general consensus.

Less generally well known is the fact that two
days before the discovery date, a paper appeared on arXiv 
\cite{DDlargeangle}, not only reiterating the neutron star merger hypothesis, 
but predicting that a SHB found in combination with a GW
would be seen far off axis. The prediction was based on
the much greater red-shift reach of GRB or SHB observations relative to the GW ones
and the fact that the $\gamma$ rays are extremely collimated. Thus,
within the volume reach of GW observations, it would be most unlikely for a
SHB to point close to the observer.

Since 1997 only two theoretical models of GRBs and their afterglows (AGs) --the standard
 fireball (FB) model \cite{FBM Reviews}
  and the cannonball (CB) model \cite{DD2004}-- have been extensively used
 to interpret the innumerable observations. Advocates of both models have
 claimed to fit the data very well. But the two models were originally
 and still are quite different in their basic assumptions, despite the repeated
 replacements of key assumptions of the ``standard" FB model (but not its
 name) with assumptions underlying the CB model (e.g.~supernovae of Type Ia
 as progenitors of most GRBs, highly collimated ejecta made of ordinary matter, 
 as opposed to spherical or conical shells of an $e^+\,e^-\,\gamma$ plasma, ``jets" not necessarily
 seen almost on axis). For a recent extensive discussion of the observational tests
 of FB and CB models, see \cite{CTDDD}.
 
 Significantly, and in contrast to the FB model(s), the CB model has made many successful
 {\it predictions}. Among them, the large polarization of the GRB's $\gamma$ rays \cite{jet},
 the precise date at which the supernova associated with GRB030329 would be
 discovered {\cite{DDD030329}, the complex ``canonical" shape \cite{Dado2002}
 of many GRB afterglows \cite{Vaughan,Cusumano}, the correlations 
between various prompt\footnote{{\it Prompt} customarily refers to quantities measured
prior to the afterglow phase. Naturally, the distinction is not always sharp.}
observables amongst them \cite{CorrelsCB,Correlations}
or with AG observables \cite{Dado 2013}.
 
The SHB170817A event is an optimal case to tighten
the discussion of the comparisons between different models of GRBs. 
The question of the apparently superluminal motion of the source
of its afterglow, discussed in chapter \ref{sec:super}, is particularly
relevant.

\section{The cannonball model}

The CB model is based on a straightforward analogy of a
phenomenon that is abundantly observed but poorly
understood: the relativistic ejecta emitted by quasars and micro-quasars.
The model \cite{jet} is illustrated in Figure \ref{fig:glory}. 
In it, bipolar jets of highly relativistic ordinary-matter
plasmoids (a.k.a.~CBs) are assumed to be launched as a compact
stellar object is being born.
SNe of Type Ic (the broad-line stripped-envelope
ones) thus generate long-duration GRBs by the electrons in a CB raising the photons
in the SN's ``glory" by Inverse Compton Scattering (ICS) into a forward-collimated
narrow beam of $\gamma$-rays \cite{jet}.

Similarly, in the CB model, mergers of two neutron stars (NSs) or a NS and a black hole
(BH) give rise to SHBs. 
In this case, the role of the glory of light is played by a Pulsar Wind Nebula (PWN) 
powered by the spin-down of a newly born rapidly-rotating pulsar-- 
suggesting that most SHBs are produced by
 NS mergers yielding a NS remnant rather than a black hole \cite{Dado2018,Dai98}.
SN-less GRBs are produced in high-mass X-ray binaries, as a NS accreting mass
from a companion suffers a phase transition to a denser object. 
Finally X-ray flashes (XRFs) and some X-ray transients are simply GRBs observed from a relatively 
large angle relative to the CBs' emission axis.

\begin{figure}
\centering
\includegraphics[width=7.5 cm]{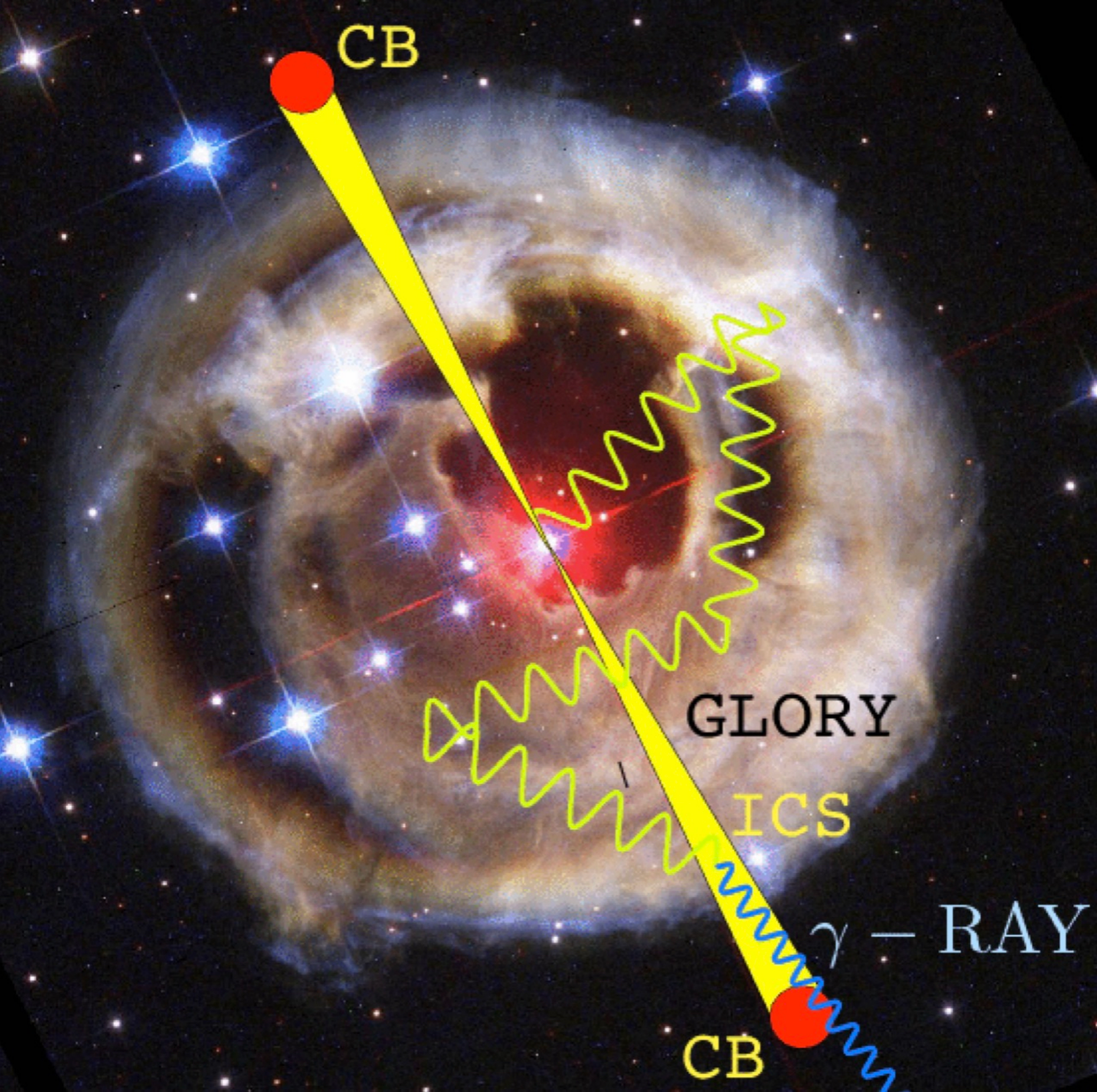}
\caption{Electrons in a cannonball inverse Compton scatter photons in the glory of light
surrounding a newly-born compact object, launching them forward as a narrow beam of 
$\gamma$ rays.}
\label{fig:glory}
\end{figure}

\section{Is SHB170817A noteworthy all by itself?}

A first question regarding SHB170817A is whether or not it is a typical SHB. The CB
model provides a strongly affirmative answer, in all respects. A first test employs the
correlations between observables predicted by this model.

Let $\gamma_0$ be the initial Lorentz factor with which a CB is launched.
Its electrons inverse-Compton-scatter the ambient photons they encounter.
This results in a $\gamma$-ray pulse of aperture $\simeq\! 1/\gamma_0\!\ll\! 1$
 around the CB's direction.
Viewed by an observer at an angle $\theta$ relative to the CB's direction, the individual photons
are boosted in energy by a Doppler factor 
$\delta_0\!\equiv\!\delta(t\!=\!0)\!=\!1/[\gamma_0\,(1\!-\!\beta_0\,\cos\theta)]$
 or, to a good
approximation for $\gamma_0^2\!\gg\!1$ and $\theta^2\!\ll\!1$, 
 $\delta_0\!\simeq\!2\gamma_0/(1\!+\!\gamma_0^2\theta^2)$.
 
The ICS of photons of energy $\epsilon$ by a CB boosts their energy, as seen by an
observer at redshift $z$, to $E_\gamma\!=\!\gamma_0\,\delta_0\,\epsilon/(1\!+\!z)$. 
Consequently, the peak energy, $E_p$, of their time-integrated energy distribution satisfies
\begin{equation}
(1+z)\,E_p\!\approx\! \gamma_0\,\delta_0\, \epsilon_p , 
\label{eq:Ep0}
\end{equation}
with $\epsilon_p$ the characteristic or peak energy of the initial photons (for the glory
of a SN $\epsilon_p\!=\!{\cal O}(1)$ eV, for a PWN $\epsilon_p\!\sim\!{\cal O}(1)$ keV).

In the Thomson regime the nearly isotropic distribution (in the CB's rest frame)
of a total number $n_\gamma$
of IC-scattered photons is beamed into an angular distribution
$dn_\gamma/d\Omega\!\approx\! (n_\gamma/4\,\pi)\,\delta^2$
in the observer's frame. Consequently, the isotropic-equivalent
total energy of the photons satisfies
\begin{equation}
E_{iso}\!\propto\! \gamma_0\, \delta_0^3\, \epsilon_p. 
\label{eq:Eiso}
\end{equation}
Hence, both ordinary long- and short-duration GRBs, viewed
most probably from an angle $\theta\!\approx\!1/\gamma$ 
(for which $\delta_0\!\approx\!\gamma_0$), are predicted \cite{CorrelsCB} to satisfy 
the ``Amati" correlation \cite{Amati2002}, 
\begin{equation}
(1+z)\,E_p\propto [E_{iso}]^{1/2},
\label{eq:Corr1}
\end{equation}
shown in Figure \ref{fig:epeiso17GRBs}.

\begin{figure}[]
\centering
\includegraphics[width=8.5 cm]{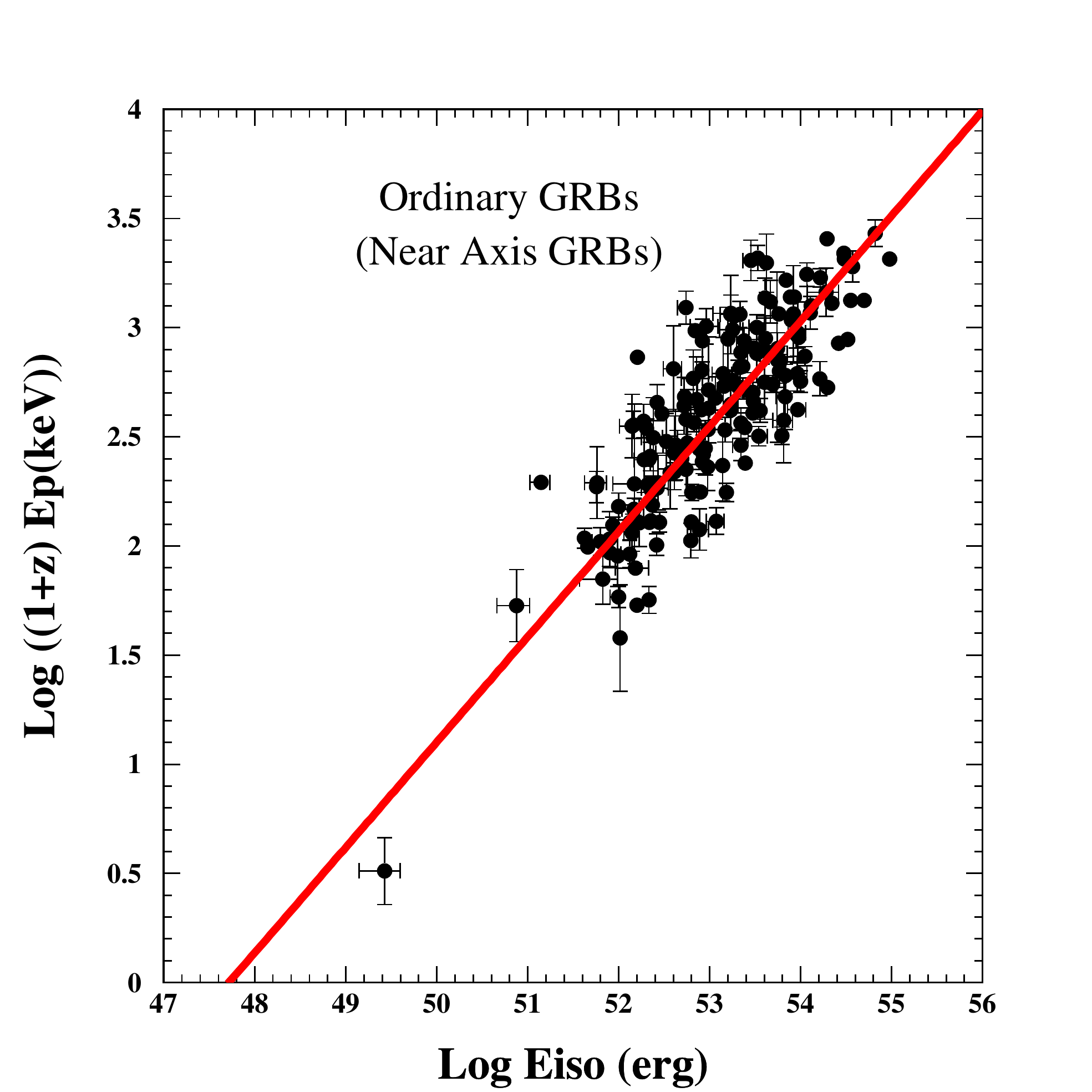}
\caption{The $[E_p, E_{iso}]$ correlation in ordinary LGRBs viewed near axis. 
The line is the best fit, whose slope, $0.48\!\pm\! 0.02$,
agrees with the CB model prediction of 
Equation \ref{eq:Corr1}. 
The lowest $E_{iso}$   GRB  is  020903 at $z\!=\!0.25$  (HETE).}
\label{fig:epeiso17GRBs}
\end{figure}

Far off-axis GRBs [$\theta^2\! \gg \! 1/\gamma^2$, so that 
$\delta_0\!\approx\! 2/(\gamma_0\,\theta^2)$],
have a much lower $E_{iso}$, and satisfy 
\begin{equation}
(1+z)\,E_p\propto [E_{iso}]^{1/3}.
\label{eq:Corr2}
\end{equation}
As one can see in Figure \ref{fig:epeisoallshbs}, SHB170817A ``is right where it should be".
As we shall discuss anon, it was indeed seen far off-axis.

\begin{figure}[]
\includegraphics[width=8.5 cm]{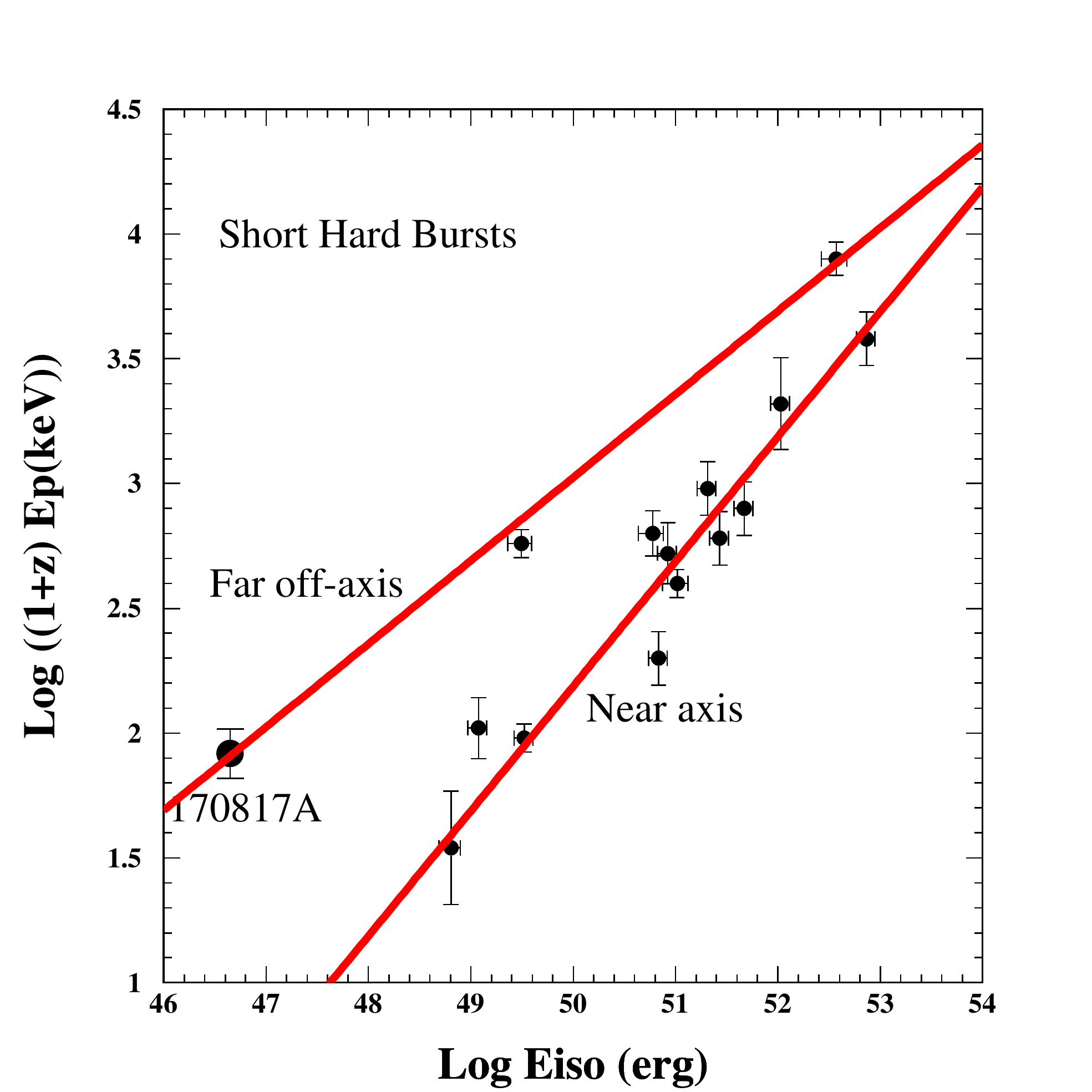}
\caption{The $[E_p,E_{iso}]$ correlations in SHBs.
The lines are the CB model predicted correlations as 
given by Equations \ref{eq:Corr1} and \ref{eq:Corr2}.}
\label{fig:epeisoallshbs}
\end{figure}

In the CB model the peak-time of a single $\gamma$-ray
pulse obeys $T_p\!\propto\!(1\!+\!z)/\gamma_0\, \delta_0$,  
and its peak energy, $E_p$, is that of Equation \ref{eq:Ep0}.
SHB170817A,
being a one-peak event, is a good case to study these observables. Indeed, one
of the simplest CB-model predictions is the $[E_p,T_p]$ correlation $E_p\!\propto\! 1/T_p$.
In Figure \ref{fig:epvt_shb} this correlation is compared 
with the values of $E_p$ and $T_p$ in the GCN circulars 
\cite{GCN}
for resolved SGRB pulses. 
Again, SHB170817A is where it should be.

\begin{figure}[]
\centering
\includegraphics[width=8.5 cm]{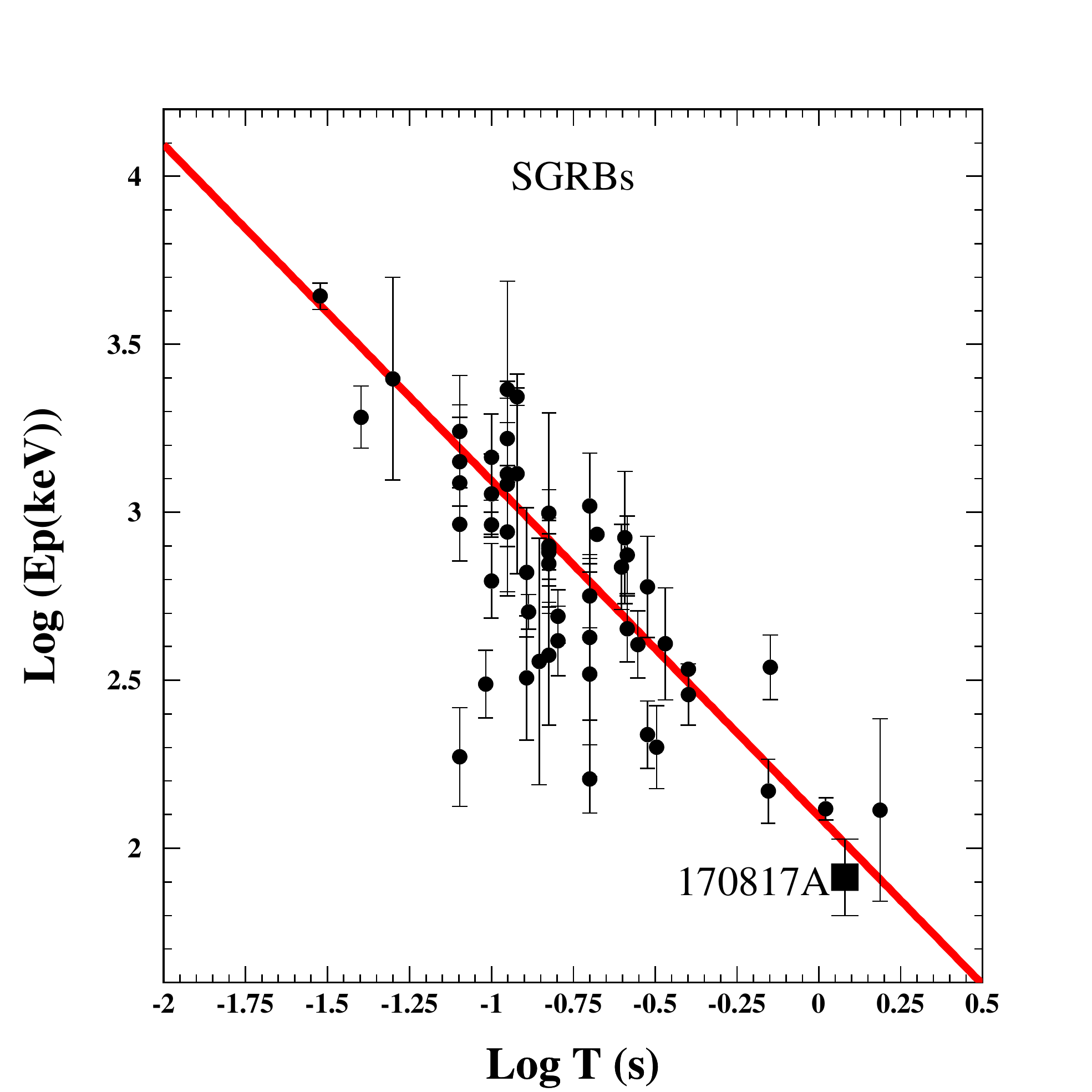}
\caption{
Comparison between the predicted  correlation ($E_p\!\propto\! 1/T_p$)
 and the corresponding data in GCN circulars
for 54 resolved pulses of SGRBs, obtained by the Konus-Wind
and Fermi-GBM collaborations.}
\label{fig:epvt_shb}
\end{figure}  

A more detailed test concerning the run-of-the-mill nature of SHB170817A is
the shape of its single pulse of $\gamma$'s. 
The light ``reservoir" that a CB will Compton up-scatter (a SN glory or a PWN) has a 
thin thermal bremsstrahlung spectrum and a number density distribution decreasing
with distance to the CB source as $1/r^2$ \cite{ThBrem}. With these inputs it is 
straightforward to derive a two-parameter simple expression that provides excellent
descriptions of pulse shapes \cite{Dado2009a}. 
A particularly well measured GRB pulse is shown in
Figure \ref{fig:Pulse930612}. 
\begin{figure}[]
\centering
\includegraphics[width=9.cm]{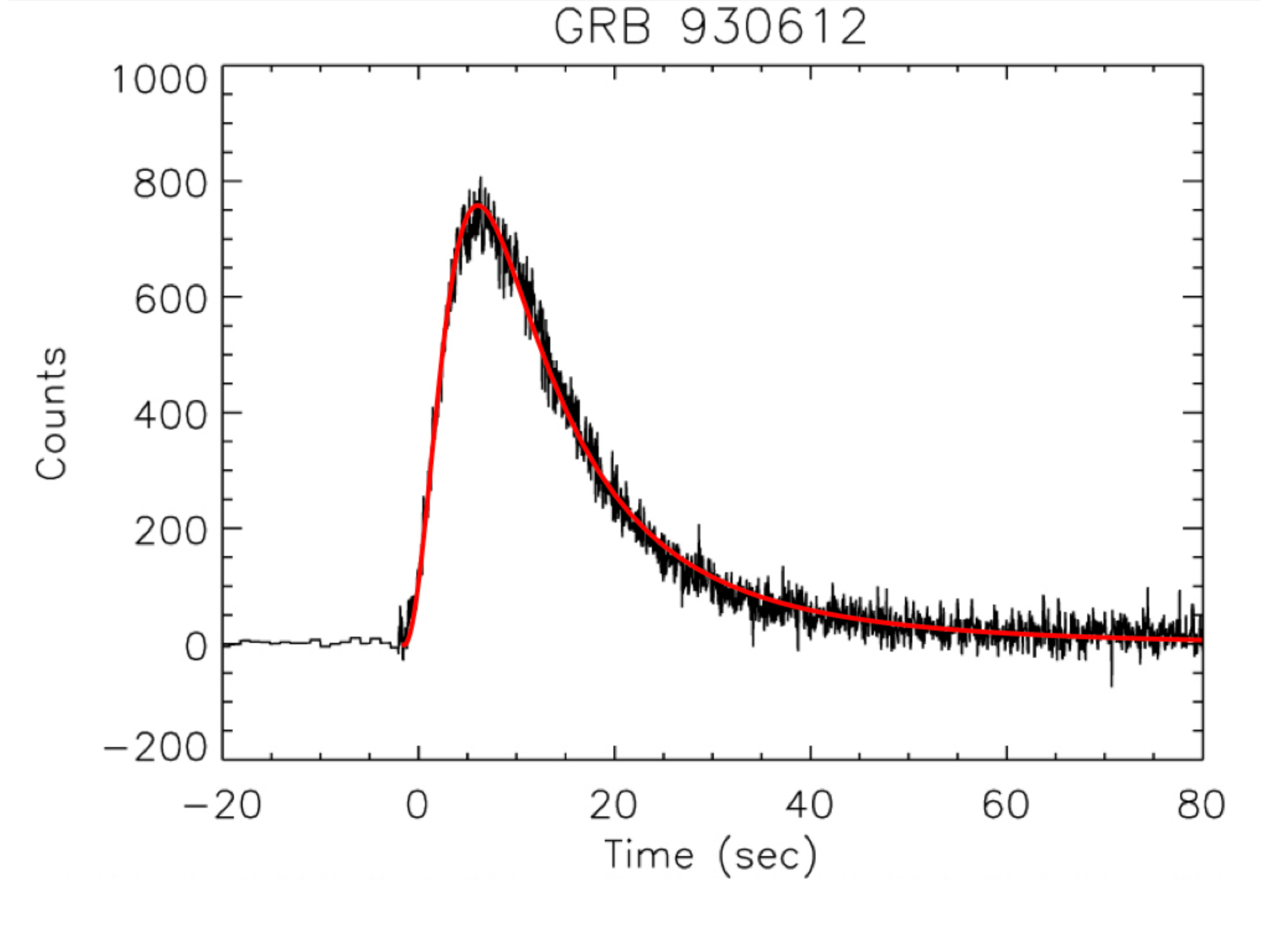}
\vspace{-1cm}
\caption{The pulse shape of GRB930612 measured with BATSE 
 aboard CGRO and 
the best CB-model fit to re-bined data \cite{DDD2022}}
\label{fig:Pulse930612}
\end{figure}

The pulse of SHB170817A is shown in Figure \ref{fig:fig02}. 
Once more, there is nothing atypical about it.
\begin{figure}[]
\centering
\includegraphics[width=8.5 cm]{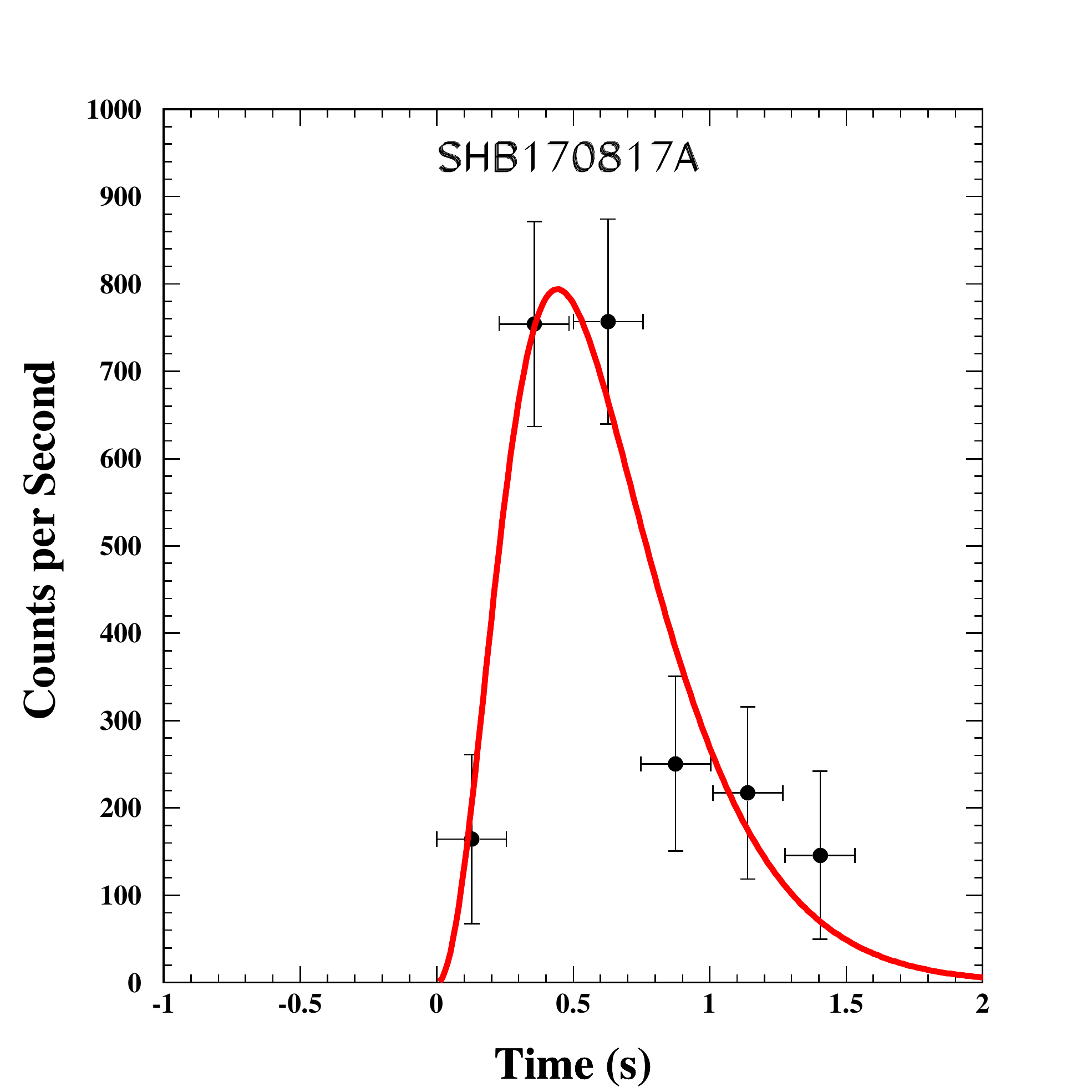}
\caption{The pulse shape of SHB170817A measured with the Fermi-GBM 
 \cite{20a,20c,20d,20e,20b} and its CB-model best fit, with 
 $\chi^2/\rm {dof}\!=\!0.95$
 \cite{DDD2022}. 
}
\label{fig:fig02}
\end{figure}

\subsection{Prompt observables of SHB170817A. Further checks}
\label{subsec:chesks}

{\it Initial Lorentz factor.}
In the CB model GRBs and SHBs can be understood as approximately standard 
candles viewed from different angles. 
 As illustrated in Figure \ref{fig:epeisoallshbs}, low luminosity (LL) 
SHBs such as SHB170817A are ordinary (O) SHBs, but viewed far off-axis.
Consequently, their $E_{iso}$ and $E_p$ are expected to obey the relations
\begin{equation}
E_{iso}{\rm (LL\,SHB)}\! \approx\! \langle\! E_{iso}{\rm (O\,SHB)}\!\rangle /
[\gamma^2\,(1\!-\!\cos\theta)]^3 \,,
\label{eq:LLSHBEiso}
\end{equation}
\begin{equation}
(1+z)\,E_p{\rm(LL\,SHB)}\!\approx\! \langle (1+z)\,E_p{\rm (O\,SHB)}\rangle /
[\gamma^2\,(1\!-\!\cos\theta)].
\label{eq:LLSHBEp}
\end{equation} 

Given the measured value  $E_{iso}\!\approx\! 5.4\times 10^{46}$ erg of 
SHB170817A \cite{Goldsteinetal2017}, the mean 
value $\langle E_{iso}\rangle\!\approx\!1\times 10^{51}$ erg of 
ordinary SGRBs,  and  the  viewing angle 
$\theta\!\approx\!28$ deg obtained --as we shall discuss in detail in section \ref{sec:superluminal}-- 
from the superluminal motion of the source of 
the radio AG of SHB170817A \cite{MooleyJET}, Equation \ref{eq:LLSHBEp}  
yields  $\gamma_0\!\approx\!14.7$ and $\gamma_0\,\theta\!\approx\! 7.2$.

With $\gamma_0$ and $\theta$ specified,
Equations \ref{eq:LLSHBEiso} and \ref{eq:LLSHBEp}, 
result in the following additional tests of CB model predictions:

{\it Peak energy.}
Assuming that SHBs have the same redshift distribution as GRBs
(with a mean value $\langle z\rangle \!\approx\!2$), and given the observed 
$\langle E_p\rangle\!=\!650$ keV of SHBs \cite{Goldsteinetal2017},
one obtains $\langle(1\!+\!z)\,E_p\rangle\! \approx\! 1950$ keV. 
Consequently, Equation \ref{eq:LLSHBEp} with $\gamma_0\,\theta\!\approx\! 7.2$ and
$z\!\approx\!1$
yields  $E_p\!\approx\!75$ keV for SHB170817A. This is to be
compared with $E_p\!=\!82 \pm 23$ keV ($T_{90}$) reported in
\cite{20a},  
$E_p\!=\!185 \pm 65$ keV estimated in \cite{20c},
and  $E_p\!\approx\!65\!+\!35(\!-\!14)$ keV estimated in \cite{20d}, 
from the same data, with a mean value $E_p\!=\!86\!\pm\!19$ keV,
agreeing with the expectation.

{\it Peak time.}
In the CB model the peak time $\Delta t$ after the beginning 
of a GRB or SHB pulse is roughly equal to half of its 
full width at half maximum
(FWHM) \cite{Dado2009a}.
Assuming again that SHBs are roughly standard candles, the
dependence of their $\Delta t$ values on $\theta$ is 
\begin{equation}
\Delta t{\rm (LL\,SHB)}\,\approx\! \gamma_0^2\,(1\!-\!\cos\theta)
\langle \Delta t{\rm (O\,SHB)}\rangle,
\label{eq:Peakt}
\end{equation}
For  $\theta\!\approx\!28$ deg,
$\Delta t\!\approx\! 0.58$ s obtained  from the prompt emission pulse 
of  SHB170817A (see Figure \ref{fig:fig02}), and $\langle{\rm FWHM(SHB)}\rangle\!=55$ ms, 
Equation \ref{eq:Vapp2} results in $\gamma_0\!\approx\! 14.7 $. 
 Using Equations 
\ref{eq:LLSHBEiso} and \ref{eq:LLSHBEp}, and $\gamma_0\,\theta\!\simeq \!7.2$ 
one checks that this value of
$\gamma_0$ is  consistent 
with  $E_{iso}\!=\!5.4\times 10^{46}$ estimated in \cite{Goldstein2017}, and 
$\langle E_p\rangle\!=\! 86\!\pm\! 19$ keV  
the mean of the estimates in  \cite{Dado2018}.

In the CB model the shape of resolved SHB and GRB  pulses 
satisfies $2\,\Delta t\!\approx\! {\rm FWHM} \!\propto\! 1/E_p$, as illustrated in
Figure \ref{fig:epvt_shb}. Using the observed 
$\langle {\rm FWHM(SHB)}\rangle\!\approx\!55$ ms, $\gamma_0\!\sim\!14.7$, 
and  $\theta\!\approx\! 28$ deg, Equation \ref{eq:Peakt} for SHB170817A results in
$\Delta t \!\approx\!0.63$ s, in good agreement with its observed value, 
$0.58\!\pm\! 0.06$ s. 

Quite obviously, the replacements we have been making of physical 
parameters by their means may not be completely reliable,
not only because of the spread in their values, but 
also because of detection thresholds and selection effects. Yet, in checking the
unexceptional nature of SHB170817A, they work better than one could expect.

\section{SHB170817A in FB models, prompt observables}

The correlations in Figures \ref{fig:epeisoallshbs},\ref{fig:epvt_shb} are trivial consequences of 
GRBs being narrow beams of $\gamma$ rays seen somewhat off-axis. They are not predictions
of FB models. In them, the jetted beams were generally
assumed to be seen on-axis, at least up to the observation
of SHB170817A. 

In FB models \cite{FBM Reviews} the GRB prompt pulses are produced by
synchrotron radiation from shock-accelerated electrons in collisions between overtaking
thin shells ejected by the central engine, or by internal shocks in the ejected conical jet.
Only for the fast decline phase of the prompt emission, and only in the limits of very
thin shells and fast cooling, falsifiable predictions have been derived
\cite{Curvature,Kobayashi,NorrisHakkila}.
They result in a power law decay
$F_\nu(t)\!\propto\!(t\!-\!t_i)^{-(\beta + 2)}\nu^{-\beta}$, where $t_i$ is the beginning time
of the decay phase, and $\beta$ is the spectral index of prompt emission.
The observed decay of the SHB170817A pulse could only be reproduced by adjusting a
beginning time of the decay and replacing the constant spectral index of the FB model
by the observed time-dependent one \cite{Curvature}.

The multitude of tests discussed in section \ref{subsec:chesks} are only based
on Compton scattering and 
relativistic kinematics for approaching objets seen significantly off-axis, like the
prompt-observable correlations we discussed. They do not belong to FB models.

\section{The afterglow of GRBs and SHBs}

In the CB model the afterglow of GRBs is due to synchrotron radiation (SR) from the
electrons in a CB that is traveling and decelerating as it interacts with the interstellar
medium (ISM), previously fully ionized by the GRB's $\gamma$ rays. The electrons radiate
in the turbulent magnetic field generated by the merging plasmas, whose energy density
is assumed to be in equilibrium with the kinetic energy brought (in the CB's rest system)
by the ISM constituents. With these inputs, the model provides an excellent and predictive
description of the temporal and spectral dependence of GRB afterglows. 

The CB-model's description of the early AGs of ``Supernova-less" GRBs and SHBs differs from
the one of ordinary (SN-generated) GRBs. It is simply the isotropic radiation
from a pulsar wind nebula (PWN), powered by a newly born rapidly-rotating pulsar, and has an
expected luminosity \cite{Dado5} satisfying
\begin{equation}
L(t,t_b)/L(t=0)\!=\!(1\!+\!t/t_b)^{-2},
\label{eq:PWN}
\end{equation}
with $t_b\!= \! P(0)/2\, \dot P(0)$, where $P(0)$ and $\dot P(0)$ are 
the pulsar's initial period and its time derivative. 
 
 The {\it universal behaviour} \cite{Dado2017} of
Equation \ref{eq:PWN} describes well
the AG of all the SN-less GRBs and SHBs with a well
sampled AG during the first few days after burst. This is demonstrated 
in Figure \ref{fig:XAGS12SHBMSP}
for the twelve SHBs \cite{Dado2018}
from the Swift XRT light curve repository \cite{Swift}
that were well sampled in the mentioned period.

\begin{figure}[]
\centering
\includegraphics[width=8.5 cm]{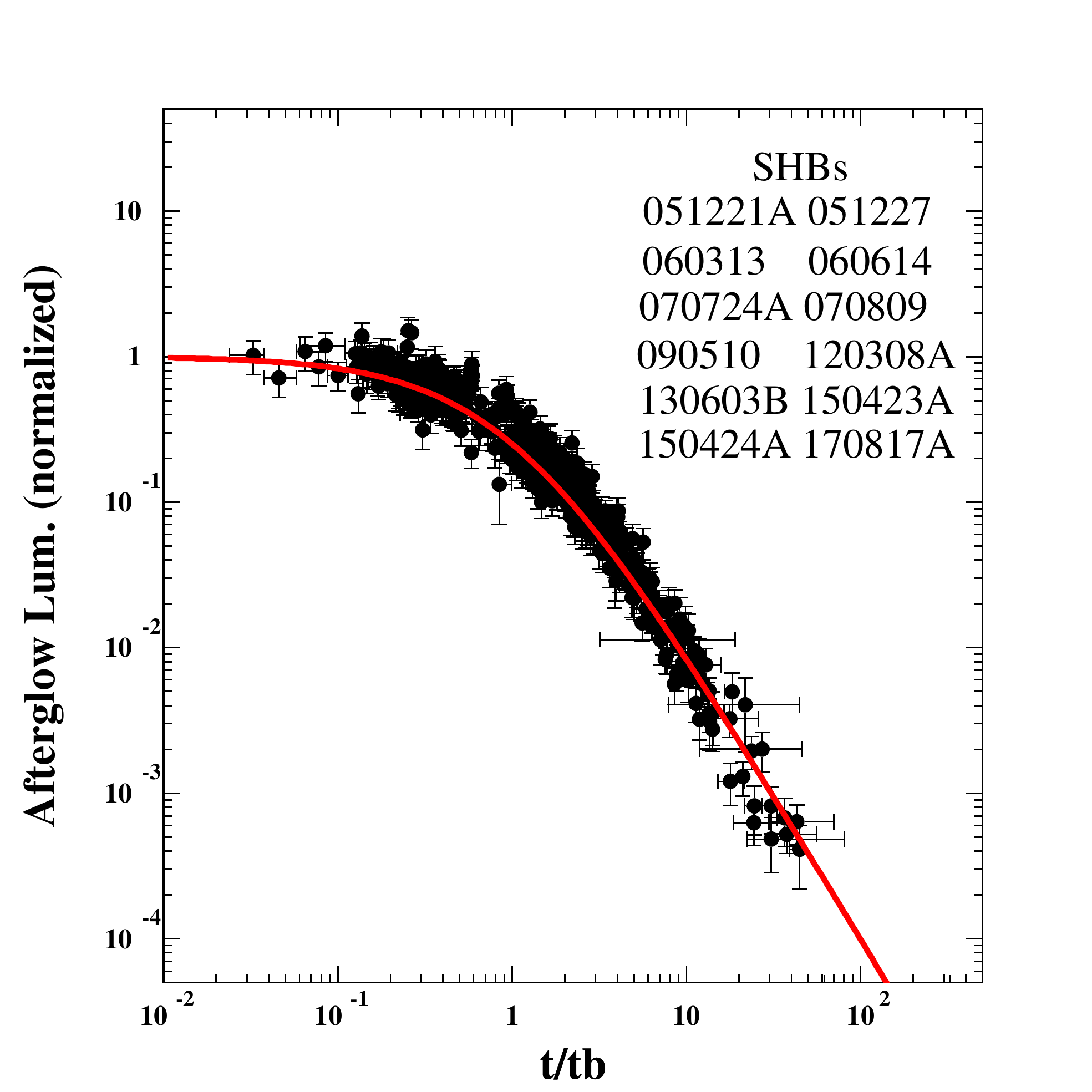}
\caption{Comparison between the normalized light curve
of the X-ray AG of 11 SHBs 
with a well sampled AG measured with Swift's XRT \cite{Swift}
during the first couple of days after burst 
and the predicted universal behavior of Equation \ref{eq:PWN}.
The bolometric light curve of SHB170817A \cite{Drout2017}
is included, and shown separately in Figure \ref{fig:fig04a}.} 
\label{fig:XAGS12SHBMSP}
\end{figure}

The bolometric light curve of SHB170817A \cite{Drout2017}
is shown in Figure \ref{fig:fig04a}. The two-parameter 
[$L(0)$ and $t_b$] CB-model fit is excellent.  SHB170817A, once more time,
is not deviant.

\begin{figure}[] 
\centering
\includegraphics[width=8.5 cm]{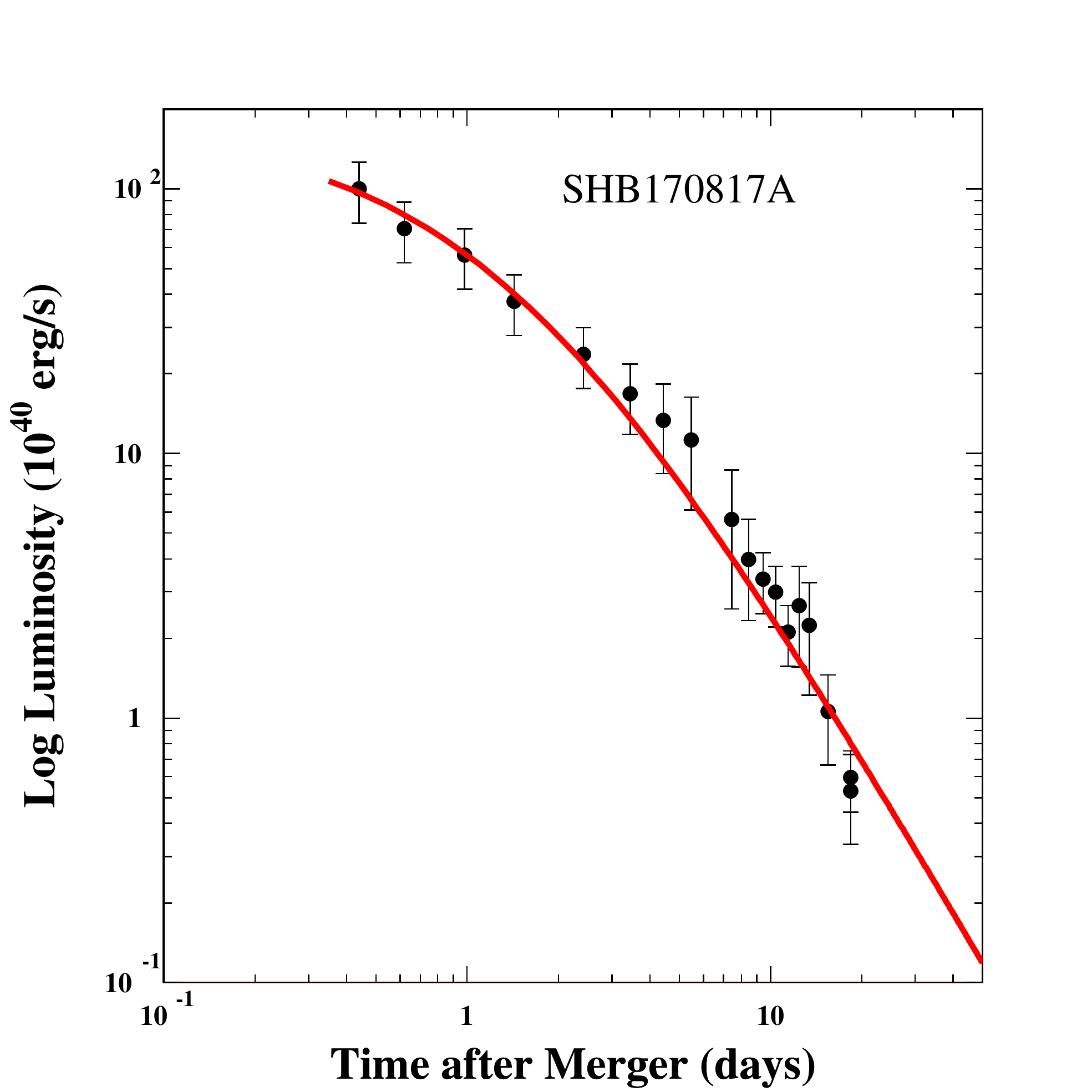} 
\caption{Comparison between the observed \cite{Drout2017} bolometric light curve 
of SHB170817A and the universal light curve of Equation \ref{eq:PWN},
assuming the presence of a milli-second pulsar  
 with $L(0)\!=\!2.27\times 10^{42}$ erg/s and  
$t_b\!=\!1.15$ d. The fit has $\chi^2/{\rm dof}\!=\!1.04$.} 
\label{fig:fig04a}
\end{figure}

\subsection{The late-time afterglow of SHB170817A}

The PWN-powered early AG decreases with time extremely fast, as in Figure \ref{fig:fig04a}. It
 is eventually overtaken, in the CB model, by the synchrotron radiation from CBs. 
In the case of SHB170817A, the AG was particularly well observed up to extremely
late times: almost three years \cite{Makha}. To discuss this subject we need to recall
some details. 

The observed spectral energy density (SED) flux of the {\it unabsorbed} SR,
$F_\nu(t)\!=\!\nu\,dN_\nu/d\nu$, has the form (see, e.g.~Eqs.~(28)-(30) in \cite{Dado2002}),
\begin{equation}
F_{\nu} \propto n(t)^{(\beta_x+1)/2}\,[\gamma(t)]^{3\,\beta_x-1}\,
[\delta(t)]^{\beta_x+3}\, \nu^{-\beta_x}\, ,
\label{eq:Fnu}
\end{equation}
where $n$ is the baryon density of the external medium encountered 
by the CB at a time $t$ and  $\beta_x$ is the spectral index 
of the emitted X-rays, $E\,dn_x/dE\propto E^{-\beta_x}$. 

The CBs are decelerated by the swept-in ionized material. Energy-momentum conservation
for such a plastic collision\footnote{The original assumption in the CB model was that the
interactions between a CB and the ISM were elastic. It was later realized, in view of the
shape of AGs at late times, that a plastic collision was  a better approximation
in the AG phase.}
--between a CB of baryon number $N_B$,  approximately constant radius $R$ \cite{DDCRs},
and initial Lorentz factor $\gamma_0\!\gg\! 1$, propagating in an
approximately constant-density ISM-- implies that the CB 
decelerates according to \cite{Dado2009a}:
\begin{equation}
\gamma(t) \simeq {\gamma_0\over \left[\sqrt{(1+\theta^2\,\gamma_0^2)^2 +t/t_d}
          - \theta^2\,\gamma_0^2\right]^{1/2}}\,,
\label{eq:decelerate}          
\end{equation}
where $t$ is the time in the observer frame since the beginning of the AG emission
by a CB, and $t_d$ is its deceleration time-scale:
\begin{equation}
t_d\!\simeq\!{(1\!+\!z)\, N_{_B}/ 8\,c\, n\,\pi\, R^2\,\gamma_0^3}.
\label{eq:td}
\end{equation}

As long as the Lorentz factor of a decelerating CB is such that $\gamma^2\!\gg\! 1$,
 $\gamma\,\delta\!\approx\!1/(1\!-\!\cos\theta)$ and
the spectral energy density of its
 synchrotron AG --Equation \ref{eq:Fnu}-- can be rewritten as 
\begin{equation}
F_{\nu}(t,\nu)\propto n(t)^{\beta_\nu+1/2}\,[\gamma(t)]^{2 \beta_\nu -4}\,\nu^{-\beta_\nu}.
\label{eq:Fnu2}
\end{equation}

For a constant density, the deceleration of the CB results in a late-time behavior
$\gamma(t)\propto t^{-1/4}$ \cite{Dado 2013}, and as long as $\gamma^2\!\gg\!1$,
\begin{equation}
F_\nu(t,\nu)\!\propto\! t^{0.72 \pm 0.03}\nu^{-0.56\!\pm\!0.06},
\label{eq:Fnu3}
\end{equation}
where we used the observed \cite{MooleySL} $\beta_\nu\!=\!0.56\!\pm\!0.06$,
which extends from the radio (R) band, through the optical (O) band, to
the X-ray band. 

If the CB moved out from within a domain of constant 
internal density into a  wind-like density
distribution (proportional to $r^{-2}$) its deceleration rate diminished
and $\gamma(t)$  became practically constant. 
Consequently, the time dependence of $F_\nu$ in
Equation \ref{eq:Fnu2} becomes a fast decline described by
\begin{equation}
F_{\nu}(t,\nu) \propto t^{-2.12\!\pm\!0.06}\nu^{-0.56\!\pm\!0.06}.
\label{eq:Fnu4}
\end{equation}

These CB-model approximate rise and fall power-law
 time dependences of the light curves of the ROX afterglow
of SHB170817, with temporal indices $0.72\!\pm\!0.03$
and $-2.12\!\pm\!0.06$, respectively, are in good agreement with 
the power-law indices extracted in \cite{MooleyKP8},
$0.78\!\pm\!0.05$  and $-2.41+0.26/-0.42$, respectively, in \cite{MooleyKP8}.
They also agree with the indices subsequently extracted
in \cite{Makha}, $0.86\!\pm\!0.04$ and $-1.92\!\pm\!0.12$, as shown
in Figure \ref{fig:RadioAG}. 

\begin{figure}[]
\centering
\includegraphics[width=9 cm]{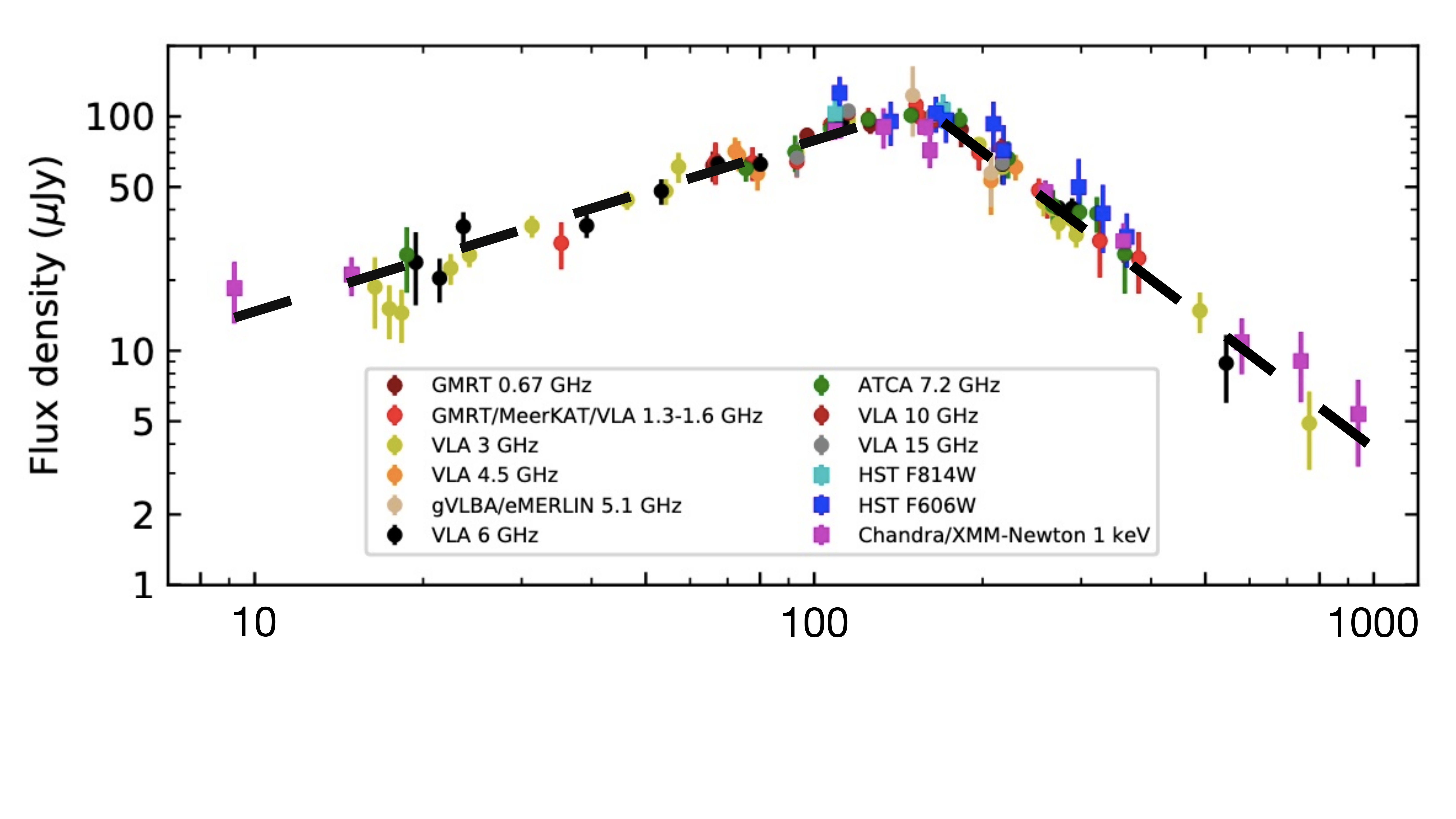} 
\vspace{-1cm}
\caption{Radio, optical and X-ray observations of the AG of SHB170817A,
adapted from Figure 2 of \cite{Makha}. The horizontal axis is the time 
after merger, in days. The radio light curve,
measured until 940 days post-merger, spans multiple
frequencies, and is scaled to 3 GHz using the spectral index $-0.584$. 
The early-time trend expected in the CB model is the rising 
black-dashed line. The late-time trend, also black-dashed, is for an assumed 
$1/r^2$ ISM density decline encountered by the CB after day $\sim\! 150$.}
\label{fig:RadioAG}
\end{figure}

Summarizing: The data in Figure  \ref{fig:RadioAG} extend
from radio to X-ray frequencies and, when corrected with the observed
spectral index, satisfactorily lie close to a single curve
(the time and frequency dependences of Equations
\ref{eq:Fnu3} and \ref{eq:Fnu4}
factorize). The dashed rising
trend is the CB-model's prediction for an assumed constant
density of the ISM encountered by the CB, which is generally an excellent
approximation. The subsequent decline follows 
from the assumption that, at an observer's time $\sim\,150$ days, 
the ISM density began to decrease as $1/r^2$. 
SHB170817A was only exceptional in that the observations lasted long
enough for this ISM-density transition to be very clearly observable.

\section{FB-model interpretations of SHB170817A} 

Soon after the discovery of the late-time radio, optical and 
X-ray afterglows of SHB170817A, many FB model best fits to the initially
rising light curves were published. They involved many different models and multiple
best-fit parameters 
(e.g.~\cite{MooleyKP7} and references therein). As new observations were made,
the proponents of FB model(s) put to use their large flexibility:

In November 2017 the authors of \cite{MooleyKP7} concluded that
{\it The off axis jet scenario as a viable explanation of 
the radio afterglow of SHB170817A is ruled out} and that a {\it chocked 
jet cocoon} is most likely the origin of the gamma rays and rising
AG of SHB170817A. 
In October 2018 the authors of \cite{MooleyKP8} reached conclusions opposite 
 to their earlier ones  \cite{MooleyKP7,Dobie2018} and to their previous
arXiv versions. To wit, in \cite{MooleyKP8} they reported a {\it strong 
jet signature in the late-time light-curve of GW170817}, and
concluded that {\it while the early-time radio emission
was powered by a wider-angle outflow (cocoon), the late-time emission was most 
likely dominated by an energetic and narrowly-collimated jet, with an opening angle of 
$<\! 5$ degrees, and observed from a viewing angle of about 20 degrees.}

All types of FB models --with conical or structured jets-- used to fit the multi-band afterglow
of SHB170817A failed to correctly predict the subsequent data.  
This is demonstrated, for example, by the arXiv versions 1-4 of 
\cite{Lazzati1} where the evolution of the AG was 
first incorrectly predicted by a structured jet with a 
relativistic, energetic core surrounded by slower and less energetic 
wings, propagating in a low density ISM, as shown in 
the upper part of Figure \ref{fig:FB_AG170817A5}. 
When the AG break around day 150 and its subsequent fast 
decline were observed, the 
structured jet model with its dozen or so adjustable parameters 
had no problem to accommodate this behavior, see 
the lower part of Figure \ref{fig:FB_AG170817A5}.

\begin{figure}[]
\centering
\includegraphics[width=6.52 cm]{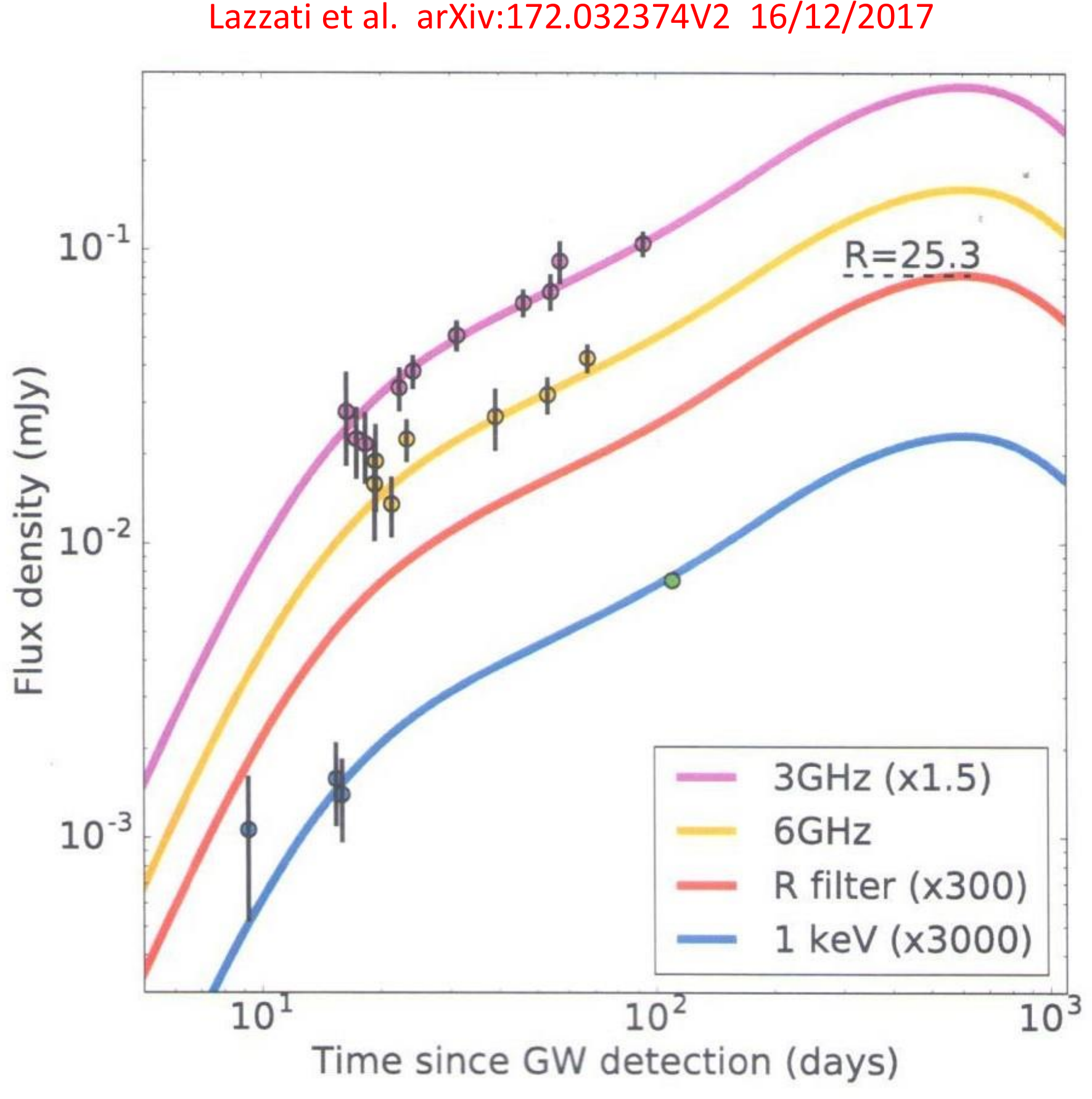} 
\includegraphics[width=7.5 cm]{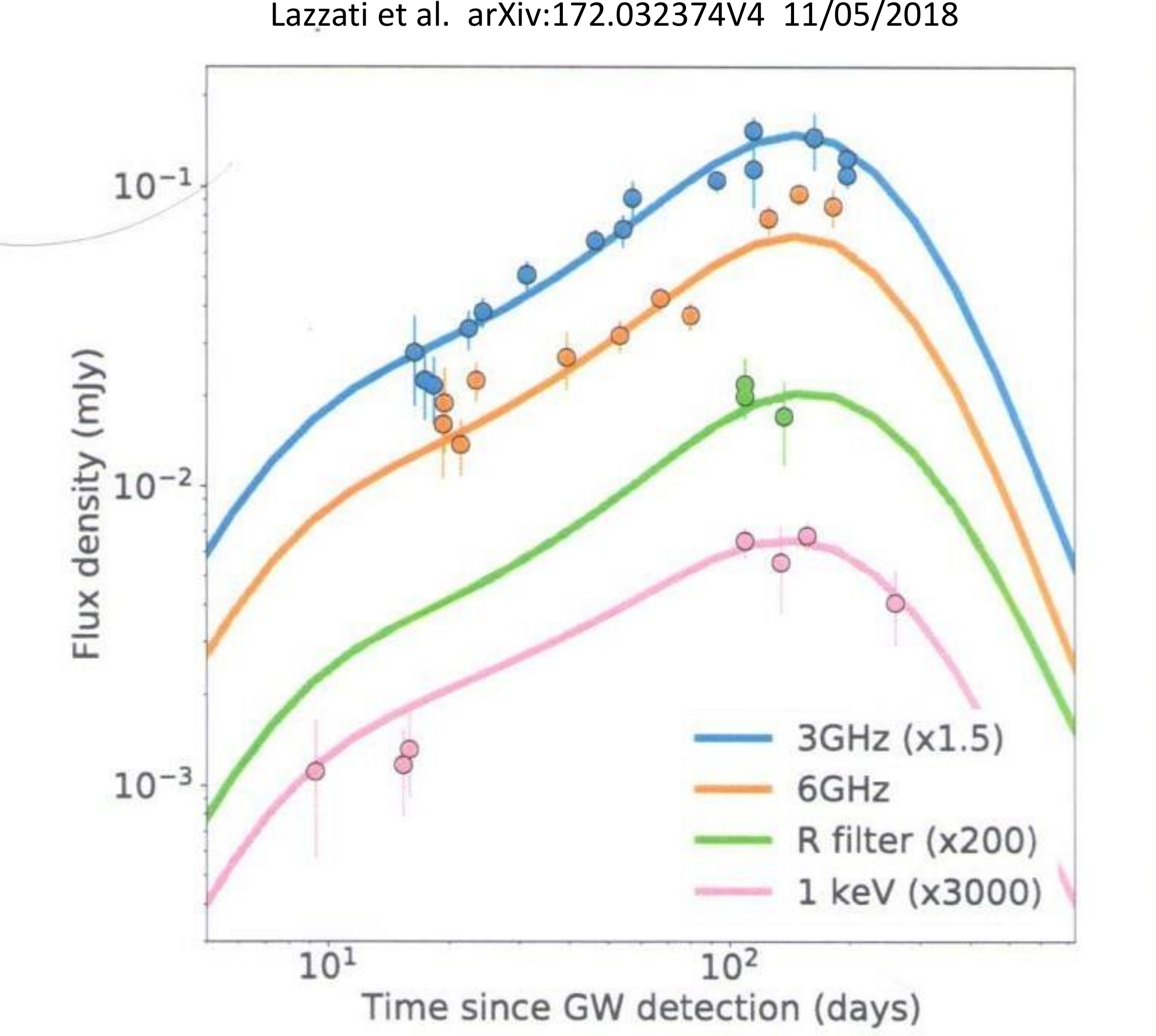} 
\caption{{\bf Above:} Best fit  light curves 
of an off-axis structured jet model 
\cite{Lazzati1} to the ROX 
AGs of SHB170817A measured before December 2017
(Figure from the first version of \cite{Lazzati1} posted in the arXiv on December 8th, 2017).
{\bf Below:} Best fit light curves to
the ROX AG of SHB170817A up to April 2018, 
obtained from a structured jet model \cite{Lazzati2}. Reported in 
version 4 of \cite{Lazzati1} posted in the arXiv on May 11th 2018.}  
\label{fig:FB_AG170817A5}
\end{figure}

In the CB model the change of slope in Figures \ref{fig:RadioAG} and
\ref{fig:FB_AG170817A5} could not be foretold, but its
a-posteriori explanation is simple. It required changing
the ISM density from a constant to a $1/r^2$ behavior, a one-parameter change.

\section{The CB's superluminal velocity in GRBs and SHBs}
\label{sec:super}

The first observation of an apparent superluminal velocity of a source in the plane of
the sky was reported \cite{Kapteyn}
 in 1902, and since 1977 in many high-resolution observations of
highly relativistic jets launched by quasars, blazars, and micro-quasars. The interpretation
of this kind of observation within the framework of special relativity was provided by Paul Courderc
in his beautiful article
{\it Les Aur\'eoles Lumineuses des Novae}
\cite{Courdec39Rees1996}.

A source with a velocity $\beta\,c$ at redshift $z$, viewed from an angle $\theta$ relative to its
direction of motion and timed by the local arrival times of its emitted photons has an
apparent velocity in the plane of the sky:
\begin{equation}
V_{app}\!=\!{\beta\,c \sin\theta \over (1\!+\!z)(1\!-\!\beta \cos\theta)}\,\approx\,
{\beta\,c\,\gamma\,\delta \sin\theta \over (1\!+\!z) }\, \cdot
\label{eq:Vapp}
\end{equation}
For $\gamma\!\gg\!1$, $V_{app}$ has a maximum value $2\,\gamma\,c/(1\!+\!z)$ at $\sin\theta\!=\!1/\gamma$.

The predicted superluminal velocity of the jetted CBs cannot be verified during
the prompt emission phase, because of its short duration and the large cosmological
distances of GRBs. But the superluminal velocity of the jet in far off-axis, i.e.~nearby
low-luminosity SHBs and GRBs, can be obtained from high resolution follow-up
 measurements of their AGs \cite{Dar2000b}. Below,  two cases are treated in detail:
 SHB170817A and GRB030329.

\subsection{The superluminally moving source of  SHB170817A}
\label{sec:superluminal}

The VLBI/VLBA observations of the 
radio AG \cite{MooleyJET} of SHB170817A provided images 
of an AG source escaping from the GRB location with
superluminal celerity.  Such a behavior in GRBs was predicted within the CB 
model \cite{DD2004} two decades ago \cite{MooleyJET}. 

Figure \ref{fig:MooleyRadio}, 
borrowed from \cite{MooleyJET}, shows the displacement with time of a compact radio source. 
The figure displays the angular locations of the radio source of the
AG of SHB170817A moving away in the plane of the
sky from the SHB location by 
$2.68\!\pm\!0.3$ mas between day 75 and day 230. 
In \cite{MooleyJET} this image is called ``a jet". It is in fact a time-lapse capture of
the moving CB emitted (approximately) towards us by the
fusion of the neutron stars.

\begin{figure}[]
\centering
\includegraphics[width=9.5 cm]{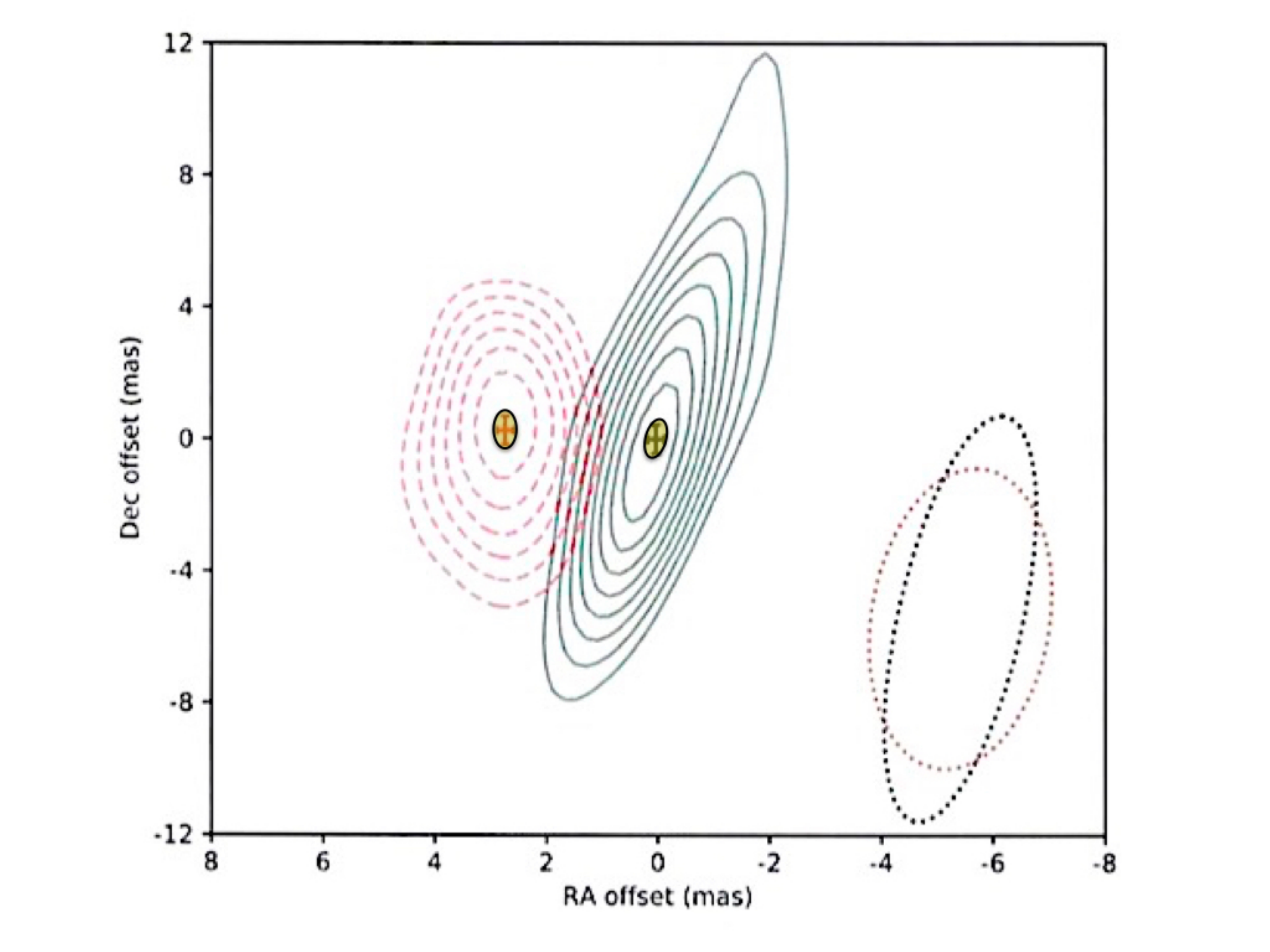}
\caption{Proper motion of the radio counterpart of GW170817. 
Its authors \cite{MooleyJET} explain: {\it The
centroid offset positions (shown by $1\,\sigma$ error bars) and 
$3\,\sigma$-$12\,\sigma$ contours of the radio source detected 75 d 
(black) and 230 d (red) post-merger with VLBI at 4.5 GHz. 
The radio source is consistent with being unresolved at both epochs. 
The shapes of the synthesized beam 
for the images from both epochs are shown as dotted ellipses in the 
lower right corner. The proper motion vector of the radio source has 
a magnitude of $2.7\!\pm\!0.3$ mas}. The $1\,\sigma$
domains have been colored
not to deemphasize the effectively point-like (unresolved) nature of the source (a CB). }
\label{fig:MooleyRadio}
\end{figure}

We have estimated that, for the CB responsible for SHB170817A,
$\gamma\!\sim\! 14.7$, that is $\beta\!\sim\!0.998$. In the excellent $\beta\!=\!1$
approximation one may rewrite
Equation \ref{eq:Vapp}, and express 
in terms of observables as:
\begin{equation} 
 V_{app}\!\approx\! 
{c\,\sin\theta\over (1\!+\!z)\,(1\!-\!\cos\theta)}\!\approx\! 
{D_A\,\Delta \theta_s\over (1\!+\!z)\Delta t}\,,
\label{eq:Vapp2}
\end{equation} 
where $\Delta\theta_s$ is the angle by which the source is seen to have moved in a time 
$\Delta t$.
The angular distance to SHB170817A to its host galaxy NGC 4993, 
at $z\!=\! 0.009783$ \cite{Hjorthetal2017}, is $D_A\!=\!39.6$ Mpc, for the local value 
 $H_0\!=\!73.4 \pm 1.62\,{\rm km/s\, Mpc}$ obtained from Type Ia SNe \cite{Riess 2016}.  
The location of the VLBI-observed source
--which moved $\Delta\theta_s\!=\! 2.70\!\pm\!0.03$ mas in a time 
 $\Delta t\!=\!155$ d (between days 75 and 230)-- implies
 $V_{app} \!\approx\! (4.0\pm 0.4)\,c$, which, solving for the
 viewing angle $\theta$
in Equation \ref{eq:Vapp2}, results in $\theta\!\approx\!  27.8 \pm 2.9$ deg.
 This value agrees with $\theta_{\rm GW}\!=\!25\pm 8$ deg, 
 the angle between the direction to the source and 
 the rotational axis of the binary system,
 obtained  from 
 the gravitational wave observations \cite{Mandeletc2017}
 for the same $H_0$ \cite{Riess 2016}. 

More strikingly, one can invert the order of the previous concordance.
If the value of $\theta_{\rm GW}$ implied by the GW observations is input in
Equation \ref{eq:Vapp2}, the result is a correct prediction of the magnitude
of the observed superluminal velocity. So simple!, this is a ``multi-messenger" 
collaboration working at its best.

\subsection{The two superluminally moving sources of GRB030329} 

This GRB was unique in the sense that it was the subject of a public 
controversy between advocates and critics of the FB and CB models. 

 As mentioned in the Introduction, the CB model was used to fit the 
 early AG of GRB030329 and to predict the discovery date of its associated
 SN, SN2003dh. This being a two-pulse GRB, the fits to its $\gamma$ rays and to its
 two-shoulder AG curve consequently involved two cannonballs. The prediction of the amount of
 their superluminal motions, based on the approximation of a constant ISM density,
 turned out to be wrong \cite{Taylor2003}. Subsequent observations of the AG showed a series of 
very clear re-brightenings, interpreted in the CB model as encounters of the CBs with ISM 
over-densities \cite{ManyBumpAG}. Corrected by the consequent faster slow-down 
of the CBs' motion, the new CB-model results were not a prediction, but
were not wrong  (see \cite{SL030329} and its Figure 2 for many details not mentioned here).

The authors of \cite{Taylor2003}
analized their data in terms of a single radio source, in spite of the fact that, with a
significance of $20\,\sigma$,
they saw two: {\it Much less easy
to explain is the single observation 52 days after the burst of an additional radio 
component 0.28 mas northeast of the main afterglow.} 
Whether there was one or two sources of the AG is the crux of the clash between
FB- and CB-model interpretations \cite{SL030329}.

Another critique by G.~B.~Taylor et al.~\cite{Taylor2003} was:
{\it
 A more general problem for the cannonball model is the absence of rapid fluctuations
in the radio light curves of GRB 030329 \cite{Berger}}. That would be true for a sufficiently
point-like source, but for CBs it is not correct. For the sake of definiteness, we discuss
the issue for this particular GRB:
 
Initially \cite{DDCRs}, the radius of a CB in its rest frame is assumed to
increase at the speed of sound in a relativistic plasma, $c_s\!=\!c/\sqrt{3}$.
At an early observer's time $T$ its radius, $R$, has increased to: 
\begin{equation}
R(T,\theta)\!\approx\! {c_s\over (1+z)}\int_0^T\delta(t,\theta)\,dt,
\label{eq:CBradius}
\end{equation}
where use has been
made of the relation between $T$ and the time in the CB's rest system. 
When the first radio observations started as early as $T\!=\!2.7$ days after burst,
the result of Equation \ref{eq:CBradius}
for this GRB --for the parameters of the CB-model description of its AG and the
deceleration law of Equation \ref{eq:decelerate}-- is $R(T)\!\approx\!5.7\!\times\! 10^{17}$ cm.
This is more than an order of magnitude larger than the largest source size 
that could still have resulted in diffractive scintillations. Case closed.

To summarize the FB advocates' views on this GRB: 
Quite forcefully Bloom and collaborators \cite{Bloom2003}
stated: {\it 
Very Long Baseline Array
imaging of the compact afterglow was used by Frail (2003) \cite{Taylor2003}
to unequivocally disprove the
cannonball model for the origin of GRBs.} On the other hand,
referring to their ``second source" the authors of \cite{Taylor2003} admitted:
{\it This component requires a high average velocity of 19c and cannot be readily 
explained by any of the standard models. Since it is only seen at a
single frequency, it is remotely possible that this image is an artifact of the calibration.}

As for cannonballs, will the dictum {\it seeing is believing}  be rejected,  
with the time-lapse image of a CB in Figure \ref{fig:MooleyRadio} somehow serving 
to disprove the cannonball hypothesis ?"

\section{Conclusion}

The cannonball model provides a successful and very simple and consistent
 interpretation of SHB170817A.
A serious limitation of the model is that the emission of relativistic
blobs of matter in core-collapse supernovae and binary-neutron-star mergers is not
theoretically well understood. Neither is, in detail, the fate of a CB traveling in the interstellar
medium. This is not surprising, for general-relativistic catastrophic 
magneto-hydrodynamics is not a simple discipline. Hopefully the renewed
interest on neutron stars and their mergers provoked by GW170817 will lead
to reinvigorated efforts on these subjects.

\section*{Acknowledgement}
I am indebted to S.~Dado,
S.~L.~Glashow and F.~Truc for their critical reading of the manuscript.
This project has received funding/support from the European Union's 
Horizon 2020 research and innovation programme under the Marie 
Sklodowska-Curie grant agreement No 860881-HIDDeN.

\end{document}